\documentclass[preprint,12pt, 3p]{elsarticle}
\usepackage{rotating}
\usepackage{graphicx}
\usepackage{booktabs}
\usepackage{amsmath}
\usepackage{amssymb}
\usepackage{adjustbox}
\usepackage{multirow}
\usepackage{subcaption}
\usepackage{xcolor}
\usepackage{float}
\usepackage{enumitem}
\usepackage{pdflscape}

\usepackage{pdflscape}
\usepackage{rotating}
\usepackage[normalem]{ulem}
\usepackage{dblfloatfix}

\useunder{\uline}{\ul}{}
\AtBeginDocument{%
  }

\newcommand{\AB}[1]{{#1}}

\definecolor{lightred}{RGB}{255,153,153}
\definecolor{darkred}{RGB}{139,0,0}
\definecolor{lightgreen}{RGB}{144,238,144}
\definecolor{darkgreen}{RGB}{0,100,0}
\definecolor{lightblue}{RGB}{135,206,250}
\definecolor{darkblue}{RGB}{0,0,139}
\definecolor{brown}{RGB}{165,42,42}

\begin{document}
\begin{frontmatter}

\title{MoR4Rec: Mixture-of-Recursion for Sequential Recommendation}
    \author[1]{Tiwari Ankit Satyaprakash}
    \ead{tiwariankitphd25@cs.du.ac.in}

    \author[2]{Adamya Shyam}
    \ead{adamya.shyam@krmangalam.edu.in}
    
    \author[1]{Vikas Kumar\corref{cor1}}
    \ead{vikas@cs.du.ac.in}
    
    \author[1]{Bharti Rana}
    \ead{bharti@cs.du.ac.in}
    
    \author[3]{Venkateswara Rao Kagita}
    \ead{venkat.kagita@nitw.ac.in}

    \address[1]{University of Delhi, India}
    \address[2]{K.R. Mangalam University, India}
    \address[3]{National Institute of Technology Warangal, India}
    
    \cortext[cor1]{Corresponding author}
    
\begin{abstract}
    Sequential recommender systems aim to predict a user’s next interaction by modeling temporal dependencies in historical behavior sequences. To capture such dependencies, recent approaches increasingly rely on attention-based architectures, particularly transformers. While effective at modeling long-range sequential patterns, these models apply a fixed computation depth to all interaction items, treating them as equally complex. However, user behavior is inherently heterogeneous, with many interactions following simple repetitive patterns, whereas others require modeling higher-order dependencies. This mismatch leads to inefficient utilization of model capacity and unnecessary computational overhead, particularly in deep transformer-based architectures. To address this issue, we propose \textit{MoR4Rec}, a sequential recommendation framework based on a Mixture-of-Recursion paradigm, which enables item-wise adaptive computation through a shared transformer backbone. A learned routing mechanism dynamically assigns each item to an appropriate recursion depth, allocating deeper computation to complex interactions while processing simpler patterns more efficiently. 
    To improve computational efficiency, we develop a cascade-based strategy with key-value sharing across recursion steps. We further introduce a gated attention mechanism that adaptively regulates contextual information during recursive sequence modeling.
    Extensive experiments on three benchmark datasets show that \textit{MoR4Rec} consistently outperforms strong transformer-based baselines, achieving average gains of 2.28\% in Recall@5, 2.50\% in NDCG@5, and 2.0\% in MRR@5 over the strongest baseline. The improvements remain consistent across top-10 and top-20 recommendation cutoffs, while comparisons with vanilla transformers demonstrate the computational efficiency of adaptive recursive processing.    

    %To address this, we propose \textit{MoR4Rec}, a novel sequential recommendation framework that introduces adaptive computation through a Mixture-of-Recursion paradigm. Unlike conventional expert-based or multi-branch models, \textit{MoR4Rec} formulates recursive computation as a shared, depth-adaptive process, where for each item, the required level of reasoning is dynamically determined. A learned routing mechanism assigns items to different recursion depths within a shared transformer backbone, enabling fine-grained specialization without explicit parameter partitioning. This design allows the model to allocate deeper computation only to complex interactions, while efficiently handling simpler interactions with shallow processing. 
    %To support this adaptive processing of items, we develop a cascade-based computation strategy with efficient key-value sharing across recursion steps, reducing redundant memory and computation. Furthermore, we introduce a gated attention mechanism that adaptively regulates contextual information during sequence modeling, enabling the model to selectively preserve relevant signals across recursive computations
    %Extensive experiments on three publicly available benchmark datasets demonstrate that \textit{MoR4Rec} consistently achieves competitive or superior recommendation accuracy while significantly improving computational efficiency compared to strong transformer-based baselines. 
\end{abstract}


\end{frontmatter}

\section{Introduction}

Sequential recommender systems (SRS) are essential for delivering personalized digital experiences across a wide range of platforms, such as e-commerce, travel, and social networks~\cite{boka2024survey, zhu2022sasnet}. These systems aim to predict a user’s future interactions by learning a mapping from historical behavioral logs to a probability distribution over candidate items. However, learning such a mapping is inherently challenging because user preferences are heterogeneous and non-stationary. A single interaction sequence often reflects multiple underlying behavioral dynamics, including short-term intent reflecting immediate needs, recurring consumption patterns shaped by habitual preferences, and long-term interests that evolve gradually over time. To model such behavioral diversity, attention-based architectures have become a primary focus in sequential recommendation research~\cite{liu2023linrec, yuan2020attention}. Among these, transformer architectures have emerged as the de facto standard due to their ability to model complex, long-range dependencies in user interaction sequences~\cite{kang2018self, sun2019bert4rec, du2023frequency, shin2024attentive,10.1145/3336191.3371786}. 

Transformer-based models usually process histories as time-ordered item embeddings, which are then processed through multiple encoding layers, where dependencies across interactions are captured via self-attention to produce contextualized representations.  Though effective, these models typically use a fixed computation depth, processing every interaction through the same stack of layers regardless of its complexity. This can be suboptimal in real-world recommendation scenarios, as user interaction sequences are often highly heterogeneous. For example, repetitive or habitual interactions may require only shallow reasoning, whereas interactions that signal shifts in user preference may benefit from deeper contextual modeling. Consequently, there is a need for models that can dynamically adapt their computational depth to the complexity of individual interactions while maintaining a unified representation framework.

%Various approaches have utilized this paradigm to develop advanced sequential recommendation models. Specifically, SASRec~\cite{kang2018self} employs a causal self-attention mechanism to model forward temporal dynamics for next-item prediction. In contrast, BERT4Rec~\cite{sun2019bert4rec} adopts a bidirectional attention mechanism with a masked item prediction objective, enabling the model to leverage both preceding and subsequent context. TiSASRec~\cite{10.1145/3336191.3371786} further incorporates temporal interval information between interactions, allowing it to capture time-aware sequential patterns beyond positional encoding. Several other works have leveraged attention-based mechanisms to enhance representation quality in SRSs~\cite{shin2024attentive, qiu2022contrastive}.

Recent research on transformer efficiency has largely focused on two complementary strategies, namely parameter sharing to reduce memory footprint ~\cite{lan2019albert} and adaptive computation to modulate processing depth~\cite{xin2020deebert}. Parameter efficiency is usually achieved through layer tying, where weights are reused across layers to reduce model size while preserving expressiveness. In parallel, adaptive computation methods, such as early exiting, allocate computation dynamically by processing simpler inputs with fewer layers. However, these paradigms have largely been developed in isolation. Recursive transformers offer a natural framework for unifying parameter sharing and adaptive computation~\cite{shen2022sliced}. Yet most existing recursive transformers keep the recursion depth fixed, as dynamic recursion often requires specialized training procedures and remains challenging to deploy efficiently. %Mixture-of-Recursions (MoR) takes a step further by enabling item-level adaptive recursion within a recursive transformer architecture~\cite{bae2025mixtureofrecursions}. Instead of assigning a fixed depth, a learned router dynamically determines how many recursive steps each item requires, allocating more computation to complex items while keeping simpler ones lightweight. 
Mixture-of-Recursions (MoR)~\cite{bae2025mixtureofrecursions} introduces adaptive computation by allowing different items to undergo different numbers of recursive transformations within a shared transformer architecture. A learned router determines the recursion depth for each item, so items that can be adequately represented with fewer transformations receive shallow computation, while items requiring more complex contextual modeling are processed through additional recursive steps.

Despite these advances, the adoption of such adaptive computation paradigms in sequential recommendation remains limited. In particular, existing models largely rely on uniform computation across interactions, failing to account for the varying complexity of user behavior sequences. To address this limitation, we propose \textit{MoR4Rec}, a novel transformer-based framework that enables item-wise adaptive computation through a Mixture-of-Recursion mechanism~\cite{bae2025mixtureofrecursions}. By dynamically adjusting the computational depth for each interaction within a shared recursive architecture, \textit{MoR4Rec} effectively captures heterogeneous user behavior while maintaining parameter efficiency.  To further improve efficiency and representation learning, the proposed framework inherits the recursion-aware key-value sharing of MoR and further introduces a gated attention mechanism that adaptively regulates contextual signals during sequence modeling. The key contributions are summarized as follows:

\begin{itemize}
    \item \AB{A novel sequential framework (\textit{MoR4Rec}) that dynamically routes each interaction to an appropriate recursion depth within a shared transformer architecture to enable item-level adaptive depth modeling.}

    \item \AB{A recursion-aware key-value (KV) sharing mechanism to reuse attention states across recursive steps, reducing the redundant computation substantially while preserving modeling capacity.}

    \item \AB{A lightweight gated mechanism that adaptively regulates the role of contextual signals to improve representation quality.}

    \item \AB{A set of auxiliary regularization objectives to ensure balanced and stable routing across recursion depths and prevent degenerate solutions.}

    \item \AB{Extensive experiments on three benchmark datasets against ten competitive baselines demonstrate that \textit{MoR4Rec} consistently achieves state-of-the-art performance.}
\end{itemize}

\iffalse
\begin{itemize}
    \item We propose \textit{MoR4Rec}, a novel framework that enables item-level adaptive depth modeling by dynamically routing each interaction to an appropriate recursion depth within a shared transformer architecture. 

    \item We leverage a recursion-aware key-value (KV) sharing mechanism inherent to MoR to reuses attention states across recursive steps, significantly reducing redundant computation while maintaining modeling capacity.

    \item We incorporate a lightweight gated attention mechanism to adaptively regulate the contribution of contextual signals, improving representation quality.

    \item We design auxiliary regularization objectives to ensure balanced and stable routing across recursion depths, preventing degenerate solutions.

    \item  Extensive experiments on three benchmark datasets against ten competitive baselines demonstrate that \textit{MoR4Rec} consistently achieves state-of-the-art performance.
\end{itemize}
\fi

The remainder of this paper is organized as follows. Section~\ref{sec:relatedWork} reviews the relevant literature on sequential recommendation. Section~\ref{sec:proposedApproach} presents the proposed \textit{MoR4Rec} framework. Section~\ref{sec:expSetup} describes the experimental setup. %, including datasets, baselines, and evaluation protocols. 
The results and analysis are discussed in Section~\ref{sec:resultsDiscuss}. Finally, Section~\ref{sec:conclusion} concludes the paper and outlines potential directions for future work.

\section{Related Work}
\label{sec:relatedWork}

Sequential recommendation has evolved significantly with the advancement of deep learning techniques. Early approaches primarily relied on feed-forward and MLP-based architectures to model user behavior patterns~\cite{cui2018mv, yan2019cosrec, gao2024smlp4rec}. These models capture feature interactions and co-occurrence relationships but lack an explicit mechanism for modeling temporal dependencies. To address this limitation, recurrent neural network (RNN)-based methods, particularly those built on LSTM and GRU architectures, were introduced to model sequential dynamics~\cite{liu2016context, donkers2017sequential}. Additionally, convolutional architectures have been employed to capture local transition patterns within interaction sequences~\cite{xu2019recurrent, yan2019cosrec}. While these methods improve temporal modeling, their inherently sequential nature limits parallelization and restricts their ability to capture long-range dependencies effectively.

Transformer-based models address these limitations by using self-attention to model global dependencies in parallel. SASRec~\cite{kang2018self} introduced unidirectional self-attention for autoregressive sequential recommendation, enabling selective attention over past interactions. BERT4Rec~\cite{sun2019bert4rec} extended this to bidirectional attention with masked item prediction for richer contextual representations. TiSASRec~\cite{10.1145/3336191.3371786} incorporated time intervals into attention to capture temporal dynamics beyond positional ordering. Subsequent work has explored representation quality and efficiency. BSARec~\cite{shin2024attentive} uses Fourier-based inductive bias to mitigate oversmoothing by disentangling low- and high-frequency components. DuoRec~\cite{qiu2022contrastive} addresses representation degeneration through dual contrastive learning with dropout augmentation and supervised  sampling. STRec~\cite{li2023strec} employs sparse attention to reduce redundant item-pair interactions, while ScaleRec~\cite{li2026multi} captures multi-scale temporal dynamics through dual attention across intra- and inter-patch granularities.

Improving computational efficiency is increasingly important in transformer-based sequential recommendation as longer interaction sequences raise computational and memory costs. LinRec~\cite{liu2023linrec} reduces self-attention complexity from quadratic to linear by reordering dot-product computation, while FMLPRec~\cite{zhou2022filter} replaces self-attention with frequency-domain filtering via Fast Fourier Transformation. LRURec~\cite{yue2024linear} employs linear recurrent units (LRU) for parallelizable, stable long-range modeling at sub-quadratic cost. LightSANs~\cite{fan2021lighter} uses low-rank decomposed self-attention to reduce parameters while preserving sequence modeling capacity. More recently, state space models, particularly Mamba-based architectures~\cite{gu2024mamba}, have emerged as efficient alternatives for long-sequence modeling. SSD4Rec~\cite{zhang2025ssd4rec} adopts structured state-space duality for efficient contextual modeling, while MaTrRec~\cite{zhang2025matrrec} combines Mamba with transformer attention in a hybrid pipeline. Other works integrate Mamba with graph encoders~\cite{zhang2025local} or develop lightweight bidirectional methods like SIGMA~\cite{liu2025sigma} for long-tail user scenarios.

Although these approaches improve efficiency and scalability, they still use fixed-depth architectures that apply uniform computation across interactions. This can incur unnecessary overhead for interactions requiring only shallow reasoning. To address this, we propose \textit{MoR4Rec}, the first framework to introduce the Mixture-of-Recursion paradigm to sequential recommendation. Its core idea is that interactions within a user sequence require different computational depths. By routing tokens across recursion depths, \textit{MoR4Rec} allocates more computation to complex interactions while processing simpler ones efficiently.

%Although these approaches improve efficiency and scalability, they still rely on fixed-depth architectures that apply uniform computation across all interactions. This can introduce substantial computational overhead, particularly for interactions requiring only shallow reasoning. To address this, we propose \textit{MoR4Rec}, the first framework to bring the Mixture-of-Recursion paradigm to sequential recommendation. The core idea is that interactions within a user sequence demand different amounts of computation. By routing tokens across recursion depths, \textit{MoR4Rec} allocates more computation to complex interactions while processing simpler ones efficiently.

\section{Proposed Approach}
\label{sec:proposedApproach}

\begin{figure}[ht]
    \centering
    \includegraphics[width=0.3\linewidth]{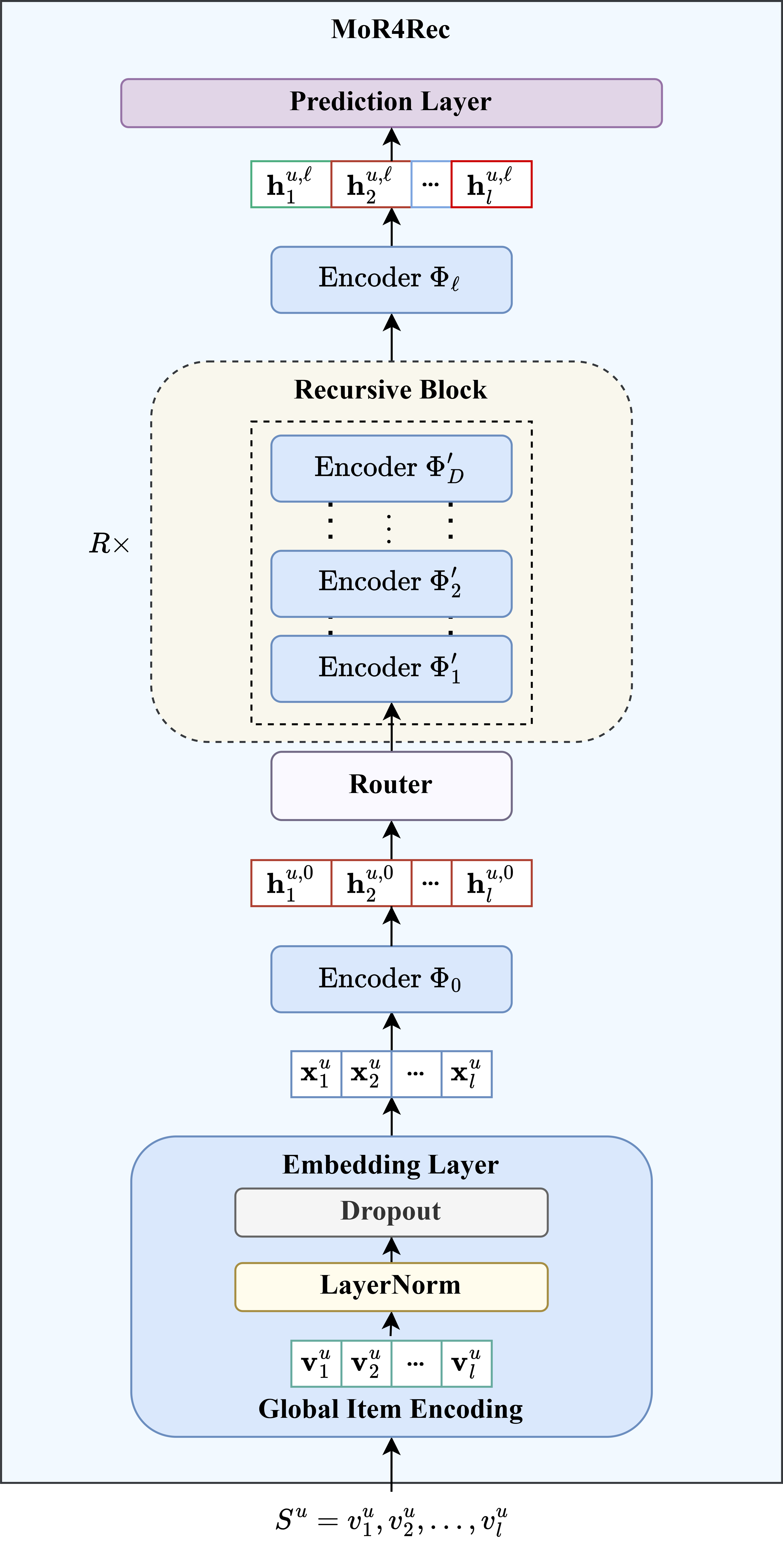}
    \caption{Overview of \textit{MoR4Rec}.}
    \label{fig:proposedApproach}
\end{figure}

In this section, we describe the proposed \textit{MoR4Rec} framework in detail. The proposed framework processes a user interaction sequence through five sequential stages as, (i)~item sequence encoding, (ii)~an initial transformer encoding layer for contextual feature extraction, (iii)~a shared recursive middle block with recursion-wise KV sharing, (iv)~a token router that assigns each item a recursion depth, and (v)~a final transformer refinement layer followed by next-item prediction. The overall pipeline of \textit{MoR4Rec} is illustrated in Figure \ref{fig:proposedApproach}. We begin by formulating the problem, followed by a detailed description of each component in the subsequent sections.

\subsection{Problem Definition}
\label{sec:problem}
 
Let $\mathcal{U} = \{u_1, \ldots, u_m\}$ denote the set of users and $\mathcal{I} = \{v_1, \ldots, v_n\}$ denote the set of items, where $m$ and $n$ represent the numbers of users and items, respectively. Each user $u \in \mathcal{U}$ has a chronologically ordered interaction history $S^u = \{v_1^u, v_2^u, \ldots, v_l^u\}$, where $v_t^u \in \mathcal{I} $ is the item interacted by user $u$ at timestep $t$ and $l$ is the sequence length. The objective of sequential recommendation is to predict the next item $v_{l+1}^u$ by maximizing a scoring function $g$ over all candidate items:
\begin{equation}
    v_{l+1}^u = \arg\max_{v \in \mathcal{I} \setminus S^u}\; g(u, S^u, v).
\end{equation}

\subsection{Item Sequence Encoding}
We begin by constructing input representations for the user interaction sequence, which serve as inputs to the subsequent recursive transformer architecture. 
Each item $v \in \mathcal{I}$ is associated with a learnable global embedding $\mathbf{v} \in \mathbb{R}^d$. Given the interaction sequence $S^u = \{v_1^u, \ldots, v_l^u\}$, input representations are formed by normalizing and regularizing each item embedding as:
\begin{equation}
    \mathbf{x}_t^u = \mathrm{Dropout}\!\left(\mathrm{LayerNorm}\!\left(\mathbf{v}_t^u\right)\right),
    \label{eq:encoding}
\end{equation}

\noindent yielding the input matrix $\mathbf{X}^u = [\mathbf{x}_1^u, \ldots, \mathbf{x}_l^u] \in \mathbb{R}^{d \times l}$. Notably, no absolute positional vectors are added at this stage. Sequential order is instead injected directly into the attention layers via Rotary Positional Embeddings (RoPE), described in the next section.

\subsection{Backbone Transformer Encoder}
\label{sec}

The backbone of \textit{MoR4Rec} is a transformer encoder composed of a recursively reusable transformer block. Each block follows a pre-normalization residual architecture~\cite{xiong2020layer}, consisting of a multi-head self-attention (MHSA) module and a position-wise SwiGLU feed-forward network (FFN)~\cite{shazeer2020glu}. Given the input representation $\mathbf{H}^{(\ell)}$ at layer index $\ell$, the transformer block is defined as

\begin{align}
    \mathbf{Z}^{(\ell)} &= \mathrm{MHSA}\!\left(\mathrm{RMSNorm}(\mathbf{H}^{(\ell)})\right) + \mathbf{H}^{(\ell)}, \label{eq:attn_res}\\
    \mathbf{H}^{(\ell+1)} &= \mathrm{FFN}\!\left(\mathrm{RMSNorm}(\mathbf{Z}^{(\ell)})\right) + \mathbf{Z}^{(\ell)}. \label{eq:ffn_res}
\end{align}

\noindent Here, $\mathbf{H}^{(\ell)},\mathbf{Z}^{(\ell)}\in\mathbb{R}^{k\times d}$ denote the hidden representations of the $k$ items at layer index $\ell$, where $d$ is the hidden dimension.

\noindent \textbf{Self-Attention with RoPE:}
At layer index $\ell$, the MHSA module operates on the normalized hidden representations
\begin{equation}
\bar{\mathbf{H}}^{(\ell)}
=
\mathrm{RMSNorm}\left(\mathbf{H}^{(\ell)}\right)
\in\mathbb{R}^{k\times d}.
\end{equation}
The query, key, and value representations are then obtained through learnable linear projections:
\begin{equation}
\mathbf{Q}^{(\ell)}
= \bar{\mathbf{H}}^{(\ell)}\mathbf{W}_Q,\qquad
\mathbf{K}^{(\ell)}
= \bar{\mathbf{H}}^{(\ell)}\mathbf{W}_K,\qquad
\mathbf{V}^{(\ell)}
= \bar{\mathbf{H}}^{(\ell)}\mathbf{W}_V,
\end{equation}
where $\mathbf{Q}^{(\ell)},\mathbf{K}^{(\ell)},\mathbf{V}^{(\ell)}
\in\mathbb{R}^{k\times d}$ are the query, key, and value matrices, respectively, and $\mathbf{W}_Q,\mathbf{W}_K,\mathbf{W}_V\in\mathbb{R}^{d\times d}$ are the corresponding learnable projection matrices. We denote the $t$-th rows of these matrices by $\mathbf{q}_t^{(\ell)}$, $\mathbf{k}_t^{(\ell)}$, and $\mathbf{v}_t^{(\ell)}$, respectively, where $\mathbf{q}_t^{(\ell)}, \mathbf{k}_t^{(\ell)}, \mathbf{v}_t^{(\ell)} \in\mathbb{R}^{d},
 t\in\{1,\ldots,k\}$.

Since absolute positional embeddings are added to hidden representations, repeatedly injecting them across recursion steps may interfere with representation reuse. We therefore adopt Rotary Positional Embeddings (RoPE)~\cite{su2024roformer}, which encodes position directly in the query and key representations without modifying the hidden states. Specifically, RoPE divides each query and key vector into $d/2$ two-dimensional sub-vectors and rotates the $i$-th sub-vector at position $t$ by $t\theta_i$, where $\theta_i = 10000^{-2i/d}$ and $i=\{0,\ldots,\frac{d}{2}-1\}$. The corresponding rotation matrix is defined as $\mathbf{R}_i(t)
=
\begin{bmatrix}
\cos(t\theta_i) & -\sin(t\theta_i)\
\sin(t\theta_i) & \cos(t\theta_i)
\end{bmatrix}$.

Applying these rotations to all two-dimensional sub-vectors gives the position-dependent query and key representations:
\begin{equation}
\tilde{\mathbf{q}}_t^{(\ell)}
=
\mathbf{R}(t)\mathbf{q}_t^{(\ell)},
\qquad
\tilde{\mathbf{k}}_t^{(\ell)}
=
\mathbf{R}(t)\mathbf{k}_t^{(\ell)},
\label{eq}
\end{equation}
where $\mathbf{R}(t)\in\mathbb{R}^{d\times d}$ is a block-diagonal rotation matrix whose $i$-th diagonal block is $\mathbf{R}_i(t)$.

The key property of RoPE is that the interaction between a query at position $t$ and a key at position $t'$ depends on their relative positional offset. Specifically,
\begin{equation}
\left(\tilde{\mathbf{q}}_t^{(\ell)}\right)^{\top}
\tilde{\mathbf{k}}_{t'}^{(\ell)}
=
\left(\mathbf{q}_t^{(\ell)}\right)^{\top}
\mathbf{R}(t)^{\top}\mathbf{R}(t')
\mathbf{k}_{t'}^{(\ell)}
=
\left(\mathbf{q}_t^{(\ell)}\right)^{\top}
\mathbf{R}(t'-t)
\mathbf{k}_{t'}^{(\ell)}.
\end{equation}
Thus, the position-aware interaction can be expressed as
\begin{equation}
\left(\tilde{\mathbf{q}}_t^{(\ell)}\right)^{\top}
\tilde{\mathbf{k}}_{t'}^{(\ell)}
=
g\left(
\mathbf{q}_t^{(\ell)},
\mathbf{k}_{t'}^{(\ell)},
t'-t
\right),
\end{equation}
where $g(\cdot)$ denotes the interaction function between the query and key representations conditioned on their relative positional offset.

The use of RoPE is naturally compatible with our KV caching strategy (Section~\ref{sec}), as the positional rotation is applied before the key representations are stored in the cache, allowing the cached keys to retain their position-specific information across recursion depths without recomputation. The rotated queries and keys are then used to compute causal self-attention. Let $N_{\mathrm{q}}$ and $N_{\mathrm{kv}}$ denote the numbers of query and key-value heads, respectively, with $N_{\mathrm{kv}}\leq N_{\mathrm{q}}$, and let $d_h=d/N_{\mathrm{q}}$ denote the per-head dimension. The attention output is computed as
$d_h = \frac{d}{N_{\mathrm{q}}}$.

For each query head, the attention output is computed as
\begin{equation}
\mathbf{A}
=
\mathrm{softmax}\left(
\frac{\tilde{\mathbf{Q}}\tilde{\mathbf{K}}^{\top}}
{\sqrt{d_h}}
+ \mathbf{M}
\right)\mathbf{V},
\label{eq}
\end{equation}
where $\mathbf{M}\in\mathbb{R}^{k\times k}$ is the causal attention mask, with $\mathbf{M}_{i,j}=-\infty$ for $j>i$ and $\mathbf{M}_{i,j}=0$ for $j\leq i$, ensuring that the item at position $i$ can attend only to itself and preceding items.  We additionally employ Grouped Query Attention (GQA)~\cite{ainslie2023gqa}, which uses fewer key-value heads than query heads to reduce the KV cache footprint during inference.

%We further employ Grouped Query Attention (GQA)~\cite{ainslie2023gqa}, which uses fewer key-value heads than query heads. Specifically, the $H_{\mathrm{q}}$ query heads are partitioned into groups, with each group sharing a single key head and value head. Consequently, only $H_{\mathrm{kv}}$ key-value heads need to be stored in the KV cache, reducing its memory footprint while retaining multiple query heads for expressive attention modeling. This design is particularly beneficial for \textit{MoR4Rec}, where the KV cache is reused across multiple recursion steps.

\iffalse

As a result, it is naturally compatible with our KV caching strategy (Section \ref{sec:kvsharing}), since rotated keys stored in the cache remain positionally valid for queries at any recursion depth without requiring recomputation. The causal self-attention output is then computed as, 
\begin{equation}
    \mathbf{A} = \mathrm{softmax}\!\left(\frac{\tilde{\mathbf{Q}}\tilde{\mathbf{K}}^\top}{\sqrt{d_h}} + \mathbf{M}\right)\mathbf{V},
    \label{eq:attention}
\end{equation}
where $d_h = d/H$ is the per-head dimension and $\mathbf{M}$ is the causal mask ($\mathbf{M}_{i,j} = -\infty$ if $j > i$, else $0$), ensuring each item attends only to past and current positions. We additionally employ Grouped Query Attention (GQA)~\cite{ainslie2023gqa}, which uses fewer key-value heads than query heads to reduce the KV cache footprint during inference.
\fi

\noindent \textbf{Feed-Forward Network:} The FFN employs a SwiGLU activation~\cite{shazeer2020glu}. Given an individual item representation $\mathbf{h}\in\mathbb{R}^{ d\times 1}$, the FFN is defined as: 
\begin{equation}
    \mathrm{FFN}(\mathbf{h}) = \left(\mathrm{SiLU}(\mathbf{W}_{\mathrm{g}} \mathbf{h}) \odot \mathbf{W}_{\mathrm{u}} \mathbf{h}\right)\mathbf{W}_{\mathrm{d}},
\end{equation}
where $\mathbf{W}_{\mathrm{g}}, \mathbf{W}_{\mathrm{u}}$ and $\mathbf{W}_{\mathrm{d}}$ are learnable weight matrices. The gating mechanism in SwiGLU allows the FFN to selectively activate features, providing expressive nonlinear transformation with a compact parameter count.

\subsection{Middle-Cycle Parameter Sharing}
\label{sec:sharing}
\textit{MoR4Rec} organizes its transformer layers into three groups: a unique input layer $\Phi_0$, a shared recursive middle block $\{\Phi'_1, \ldots, \Phi'_D\}$, and a unique output layer $\Phi_{L-1}$. Let $L$ be the total number of logical (unrolled) layers and $R$ the number of recursion steps. The base depth of the shared block is $D = (L-2)/R$, with $L-2$ divisible by $R$. The fully unrolled hidden state evolution over the stack is:

\begin{equation}
\mathbf{h}_t^{(\ell+1)} =
\begin{cases}
f\!\left(\mathbf{h}_t^{(\ell)};\, \Phi_0\right),
& \ell = 0, \\[4pt]
f\!\left(
    \mathbf{h}_t^{(\ell)};
    \Phi'_{((\ell-1) \bmod D)+1}
\right),
& 1 \le \ell \le L-2, \\[4pt]
f\!\left(\mathbf{h}_t^{(\ell)};\, \Phi_{L-1}\right),
& \ell = L-1.
\end{cases}
\label{eq:sharing}
\end{equation}
where $\mathbf{h}_t^{(\ell)}$ denotes the hidden representation of the item at position $t$ after logical layer $\ell$, $f(\cdot,;\Phi)$ denotes a transformer encoder layer parameterized by $\Phi$, and $\ell$ indexes the logical layer in the unrolled network. Parameters are shared across recursion: the middle block is instantiated once and reused at every step, so all $R$ passes read from and accumulate gradients into one physical set of parameters, keeping the budget at $\mathcal{O}((D+2)\cdot P)$ while effective depth scales with $R$. This recurrence defines the deepest path, while routing assigns each token a depth $r$ so it traverses the shared block $r+1$ times. The key property of this design is that the total parameter count is $\mathcal{O}((D+2) \cdot P)$ regardless of $R$, where $P$ is the per-layer parameter count. Thus, the model does \emph{not} grow with the number of recursion steps. The unique input and output layers serve as domain-specific adapters. $\Phi_0$ contextualizes raw item embeddings before routing, and $\Phi_{L-1}$ refines aggregated representations before prediction.

\subsection{Token-Level Routing and Adaptive Computation}
\label{sec:routing}

The central contribution of \textit{MoR4Rec} is a token-level router that assigns each item in the sequence an adaptive recursion depth, operationalizing the hypothesis that different items require different amounts of contextual processing.

\noindent \textbf{Router:}
After the unique input layer $\Phi_0$, the router takes the contextualized representation of each item and predicts a distribution over the $R$ available recursion levels. For user $u$'s interaction at sequence position $t$, let $\mathbf{h}_t^{u,0}\in\mathbb{R}^{d}$ denote its contextualized representation after the input layer. The routing distribution is computed as
\begin{equation}
\mathbf{r}_t^u
=
\mathrm{Softmax}\left(
\mathbf{W}_r^\top \mathbf{h}_t^{u,0}
\right),
\qquad
\mathbf{W}_r\in\mathbb{R}^{d\times R},
\label{eq:router}
\end{equation}
where $\mathbf{r}_t^u\in\mathbb{R}^{R}$ contains the routing probabilities over the $R$ recursion levels. The assigned recursion depth is then determined by the level with the highest routing probability:
\begin{equation}
    z_t^u = \mathrm{argmax}_{r \in \{1,\ldots,R\}}\, \mathbf{r}_{t}^u.
    \label{eq:hard_routing}
\end{equation}
The items with recursive depth 1 pass through the shared block once, while those with recursive depth $R$ pass through it $R$ times.

\noindent \textbf{Cascading Token Selection:}
At each recursion level $r\in \{1,\ldots,R\}$, only items whose assigned recursion depth is at least $r$ participate in the forward pass through the shared block. Specifically, the active set at level $r$ is defined as
\begin{equation}
\mathcal{S}^{u,r}
=
\{
v_t^u \mid r_t^u \geq r
\},
\qquad
|\mathcal{S}^{u,1}|
\geq
|\mathcal{S}^{u,2}|
\geq
\cdots
\geq
|\mathcal{S}^{u,R}|,
\label{eq:selection}
\end{equation}
where $r_t^u$ denotes the recursion depth assigned to the $t$-th item of user $u$. The active items in $\mathcal{S}^{u,r}$ are extracted from the full sequence tensor using vectorized \texttt{gather} operations, processed through the $D$ shared layers, and then written back to their original positions using \texttt{scatter}. Items whose assigned depth is exactly $r$ exit after completing this recursion pass. Their resulting representations are weighted by the corresponding router probability and accumulated with the initial representations:
\begin{equation}
\mathbf{h}_t^{u,\mathrm{final}}
=
\mathbf{h}_t^{u,0}
+
p_t^{u,r}\mathbf{h}_t^{u,r},
\qquad
\forall\, t : z_t^u = r.
\label{eq:weighted_exit}
\end{equation}
where $p_t^{u,r}$ denotes the router probability assigned to recursion level $r$ for the $t$-th item of user $u$. This residual formulation preserves a direct path from the initial representation $\mathbf{h}_t^{u,0}$ to the final output, while the differentiable weighting factor $p_t^{u,r}$ provides a learning signal to the router despite the non-differentiable depth selection.

\subsection{Adaptive KV Sharing Across Recursion Steps}
\label{sec:kvsharing}

A key observation enabling efficient recursive processing is that every non-padding item participates in the first recursion step, i.e., $\mathcal{S}^{u,1}$ contains all valid interactions in the sequence. We therefore compute and cache the rotated key and value representations during the first recursion step and reuse them across subsequent steps. For user $u$, the initial KV cache is given by
\begin{equation}
\tilde{\mathbf{K}}^{u,1}_t
=
\mathbf{R}(t)
\left(\mathbf{W}_K\mathbf{h}_t^{u,0}\right),
\qquad
\mathbf{V}^{u,1}_t
=
\mathbf{W}_V\mathbf{h}_t^{u,0},
\qquad
v_t^u\in\mathcal{S}^{u,1},
\label{eq:kv_cache}
\end{equation}
where $\tilde{\mathbf{K}}^{u,1},\mathbf{V}^{u,1}\in\mathbb{R}^{k\times d}$ represent the cached key and value matrices for the full interaction sequence of user $u$.

At each subsequent recursion level $r>1$, the cached key-value representations are selectively updated only for active items in $\mathcal{S}^{u,r}$. Specifically,
\begin{equation}
\tilde{\mathbf{K}}^{u,r}_t =
\begin{cases}
\mathbf{R}(t)\!\left(\mathbf{W}_K\mathbf{h}_t^{u,r}\right),
& v_t^u\in\mathcal{S}^{u,r}, \\[4pt]
\tilde{\mathbf{K}}^{u,r-1}_t,
& v_t^u\notin\mathcal{S}^{u,r}.
\end{cases}
\label{eq:k_update}
\end{equation}

and

\begin{equation}
\mathbf{V}^{u,r}_t =
\begin{cases}
\mathbf{W}_V\mathbf{h}_t^{u,r},
& v_t^u\in\mathcal{S}^{u,r}, \\[4pt]
\mathbf{V}^{u,r-1}_t,
& v_t^u\notin\mathcal{S}^{u,r}.
\end{cases}
\label{eq:v_update}
\end{equation}
Thus, active items progressively refine their key-value representations across recursion steps, whereas inactive items retain their most recently computed states. At recursion level $r$, queries are computed only for the $|\mathcal{S}^{u,r}|$ active items. The resulting attention output is
\begin{equation}
\mathbf{O}^{u,r}
=
\mathrm{Softmax}\left(
\frac{
\tilde{\mathbf{Q}}^{u,r}
\tilde{\mathbf{K}}^{u,r\top}
}{\sqrt{d_h}}
+
\mathbf{M}^{u,r}
\right)
\mathbf{V}^{u,r},
\label{eq:attention_recursive}
\end{equation}
%where $\tilde{\mathbf{Q}}^{u,r}\in\mathbb{R}^{|\mathcal{S}^{u,r}|\times d}$ contains the rotated queries of the active items, while $\tilde{\mathbf{K}}^{u,r},\mathbf{V}^{u,r}\in\mathbb{R}^{k\times d}$ contain the cached key and value representations for the complete sequence. The corresponding causal mask $\mathbf{M}^{u,r}\in\mathbb{R}^{|\mathcal{S}^{u,r}|\times k}$ restricts each active query at position $t$ to attend only to positions $1,\ldots,t$, including itself. Since $|\mathcal{S}^{u,r}|$ decreases monotonically with $r$, the number of query representations requiring computation becomes progressively smaller at deeper recursion levels. Meanwhile, each active item retains access to the complete sequence context through the cached $\tilde{\mathbf{K}}^{u,r}$ and $\mathbf{V}^{u,r}$, regardless of how many other items have exited. This selective KV update avoids redundant key-value computation for inactive items while preserving full contextual access for active items, thereby reducing the computational cost of recursive attention.
where $\tilde{\mathbf{Q}}^{u,r}\in\mathbb{R}^{|\mathcal{S}^{u,r}|\times d}$ contains the rotated queries of active items, while $\tilde{\mathbf{K}}^{u,r},\mathbf{V}^{u,r}\in\mathbb{R}^{k\times d}$ store cached key-value representations for the full sequence. The causal mask $\mathbf{M}^{u,r}\in\mathbb{R}^{|\mathcal{S}^{u,r}|\times k}$ restricts each active query at position $t$ to positions $1,\ldots,t$. As $|\mathcal{S}^{u,r}|$ decreases with $r$, deeper recursion levels require fewer query computations, while active items retain full sequence context through cached keys and values. This selective KV update avoids redundant computation for inactive items while preserving contextual access for active ones.
\subsection{Gated Attention Mechanism ($G_1$)}
\label{sec:gating}

Standard attention computes a weighted sum over values and immediately projects the result to the output space. To allow the model to \emph{dynamically suppress} less informative contextual signals on a per-item basis, we introduce a gating mechanism at position $G_1$, applied to the attention output $\mathbf{A}^{u,r}$ before the output projection $\mathbf{W}_O$. We study two variants that trade off granularity against parameter cost.

\noindent\textbf{Headwise Gating:} This variant appends one extra scalar dimension per attention head to the query projection, producing a gate $g_{t,h}^{u,r}\in\mathbb{R}$ for each item $v_t^u$ and head $h$ at recursion level $r$:
\begin{equation}
\mathbf{W}_Q \in \mathbb{R}^{d \times N_q(d_h+1)}, \qquad
\hat{\mathbf{A}}^{u,r}_{t,h,:}
=
\sigma\left(g_{t,h}^{u,r}\right)
\cdot
\mathbf{A}^{u,r}_{t,h,:},
\label{eq:headwise_gate}
\end{equation}
where $\mathbf{W}_Q\in\mathbb{R}^{d\times N_q(d_h+1)}$ is the query projection matrix. The extra $N_q$ dimensions, one per head, carry the gate scalars $g_{t,h}^{u,r}$, which are extracted after reshaping, and $\sigma$ denotes the sigmoid function. This enables the model to amplify or suppress the contribution of entire attention heads for each item, at the cost of $N_h$ additional output dimensions in the query projection.

\noindent\textbf{Elementwise Gating:} This variant doubles the query projection to produce a full $d$-dimensional gate vector per item, enabling feature-level modulation:
\begin{equation}
\mathbf{W}_Q \in \mathbb{R}^{d \times 2d}, \quad
\left[\tilde{\mathbf{Q}}^{u,r} \,;\, \mathbf{G}^{u,r}\right]
=
\mathbf{H}^{u,r}\mathbf{W}_Q, \quad
\hat{\mathbf{A}}^{u,r}
=
\sigma\left(\mathbf{G}^{u,r}\right)
\odot
\mathbf{A}^{u,r},
\label{eq:elementwise_gate}
\end{equation}
where $\mathbf{G}^{u,r}\in\mathbb{R}^{|\mathcal{S}_r|\times N_q\times d_h}$ is the gate tensor obtained by splitting the second half of the query projection output, $\mathbf{H}^{u,r}$ denotes the hidden states of active items at recursion level $r$, and $\odot$ denotes elementwise multiplication. Elementwise gating allows fine-grained, dimension-level control over the attention output. This strategy is particularly beneficial in the presence of noisy or incidental interactions in sparse user histories. The trade-off is a $2\times$ increase in query projection output dimensions. We evaluate these variants against vanilla attention in Section~\ref{sec:ablation}.

\subsection{Prediction and Learning Objective}
\noindent \textbf{Prediction:} Given the final fused representation $\mathbf{h}_t^{u,\text{final}} \in \mathbb{R}^d$ at timestep $t$, the preference score for a candidate item $v \in \mathcal{I} \setminus S^u$ 
is computed as:
\begin{equation}
    \hat{s}_{t+1,v}^u 
    = \left(\mathbf{h}_t^{u,\text{final}}\right)^\top \mathbf{v},
    \label{eq:score}
\end{equation}
where $\mathbf{v} \in \mathbb{R}^d$ is the learnable embedding of item $v$.

\noindent \textbf{Prediction Loss:} 
Let $\mathcal{T} = \{(u,t) \mid u \in \mathcal{U},\; v_{t+1}^u \neq \texttt{[PAD]}\}$ denote the set of valid prediction positions. For each $(u,t) \in \mathcal{T}$, the probability assigned to the ground-truth next item $v_{t+1}^u$ is:
\begin{equation}
    P(v_{t+1}^u \mid u, t)
    = \frac{       \exp\!\left(\hat{s}_{t+1,\,v_{t+1}^u}^u\right)
    }{
        \displaystyle\sum_{v \in \mathcal{I} \setminus S^u}
        \exp\!\left(\hat{s}_{t+1,v}^u\right)
    }.
    \label{eq:probScore}
\end{equation}
The cross-entropy prediction loss over all valid positions is:
\begin{equation}
    \mathcal{L}_{\mathrm{rec}}
    = -\frac{1}{|\mathcal{T}|}
    \sum_{(u,t) \in \mathcal{T}}
    \log P(v_{t+1}^u \mid u, t).
    \label{eq:rec_loss}
\end{equation}

\noindent \textbf{Load Balancing Loss:}
Since hard top-1 routing via $\arg\max$ is non-differentiable, it is susceptible to routing collapse, where all items $v_t^u$ are assigned to the same recursion level $z_t^u$, eliminating adaptive computation entirely. To prevent this, we introduce a load-balancing loss following~\cite{JMLR:v23:21-0998} that encourages uniform utilization across all $R$ recursion levels. For a batch of $B$ sequences each of length $l$, let $P_r$ be the mean routing probability assigned to level $r$ and $f_r$ the empirical fraction of items hard-routed to level $r$:

\begin{equation}
    P_r = \frac{1}{Bl}\sum_{u=1}^{B}\sum_{t=1}^{l} p_t^{u,r},
    \qquad
    f_r = \frac{1}{Bl}\sum_{u=1}^{B}\sum_{t=1}^{l}
    \mathbb{I}(z_t^u = r),
    \label{eq:balance_terms}
\end{equation}
where $p_t^{u,r}$ is the router probability for item $v_t^u$ at recursion level $r$ and $z_t^u$ is the hard routing decision. The load-balancing loss is then:
\begin{equation}
    \mathcal{L}_{\mathrm{bal}} = R \sum_{r=1}^{R} P_r \cdot f_r.
    \label{eq:balance_loss}
\end{equation}
Since $f_r$ is non-differentiable, gradients flow only through $P_r$, providing a soft signal that pushes the router toward uniform dispatch without requiring straight-through estimators.

\noindent \textbf{Total Training Objective:} 
The full training objective combines the next-item prediction loss 
with the auxiliary balancing term:
\begin{equation}
    \mathcal{L} = \mathcal{L}_{\mathrm{rec}} + \mathcal{L}_{\mathrm{bal}}.
    \label{eq:total_loss}
\end{equation}

\section{Experimental Setup}
\label{sec:expSetup}
We conduct extensive experiments to systematically evaluate \textit{MoR4Rec} against competitive baselines and ablated variants.  Specifically, we aim to answer the following research questions:

\subsection{Research Questions}
\label{sec:researchQues}
Our experiments address the following research questions:
\begin{itemize}
    \item \textbf{RQ1}: How does \textit{MoR4Rec} compare against state-of-the-art sequential recommendation models in terms of 
    recommendation accuracy across diverse datasets?  (Section~\ref{sec:main_results})
    \item \textbf{RQ2}: How do the gated attention variants affect model performance? (Section~\ref{sec:ablation})
    \item \textbf{RQ3}: What recursion depth does the router assign to each token? (Section~\ref{sec:recursion})
    \item \textbf{RQ4}: Does MoR4Rec's adaptive per-token recursion lower computational cost without adding parameters? (Section~\ref{sec:flops})
    % \item \textbf{RQ5}: How sensitive is \textit{MoR4Rec} to key hyperparameters? (Section~\ref{sec:sensitivity})
    \item \textbf{RQ5}: Are the performance gains of \textit{MoR4Rec} statistically significant across datasets? (Section~\ref{sec:stats})
\end{itemize}

\subsection{Datasets}
\label{sec:datasets}
We evaluate on three publicly available benchmark datasets from two well-established repositories, Amazon~\footnote{https://amazon-reviews-2023.github.io/} and GroupLens~\footnote{https://grouplens.org/datasets/movielens/}. From Amazon, we utilize the Video Games and Industrial \& Scientific product review datasets, representing sparse and long-tail interaction patterns. From GroupLens, we adopt the relatively dense MovieLens 1M movie rating dataset. 
Detailed statistics for each dataset are summarized in Table~\ref{tab:datastats}.

\begin{table}[ht]
    \centering
    \renewcommand{\arraystretch}{1}
    \caption{Statistics of the considered datasets.}
    \adjustbox{max width=0.8\linewidth}{
    \begin{tabular}{lcccc}
        \toprule
        \textbf{Dataset} & \textbf{\# Users} & \textbf{\# Items} & \textbf{\# Interactions} & \textbf{Sparsity (\%)} \\
        \midrule
        \textbf{ML-1M} & 6040 & 3416 & 999611 & 95.16 \\
        \textbf{Video   Games} & 94762 & 25612 & 814586 & 99.96 \\
        \textbf{Industrial   \& Scientific} & 50985 & 25848 & 412947 & 99.96 \\
        \bottomrule
    \end{tabular}
    }
    \label{tab:datastats}
\end{table}

%We apply a standard 5-core filtering strategy, retaining users and items with at least five interactions. Following~\cite{yue2024linear}, we handle variable-length user interaction sequences by fixing the maximum sequence length to $l$. Sequences longer than $l$ are segmented into multiple subsequences, while shorter sequences are left-padded with a special \texttt{[PAD]} token. In all experiments, we set $l=50$, consistent with the configurations adopted by most baseline models. For data splitting, we adopt the leave-one-out protocol~\cite{kang2018self}. The most recent interaction of each user is held out for testing, the second most recent for validation, and all remaining interactions form the training set. At test time, both training and validation interactions are available as context for predicting the test item.

We apply standard 5-core filtering, retaining users and items with at least five interactions. Following~\cite{yue2024linear}, we fix the maximum sequence length to $l$; longer sequences are split into subsequences, while shorter ones are left-padded with \texttt{[PAD]}. We set $l=50$, consistent with most baselines. For data splitting, we use the leave-one-out protocol~\cite{kang2018self}, holding out each user's most recent interaction for testing, the second most recent for validation, and the remainder for training. At test time, training and validation interactions are used as context for predicting the test item.

\noindent \textbf{Evaluation Metrics.}
We evaluate recommendation quality using three standard ranking metrics: Recall@$k$, Normalized Discounted Cumulative Gain (NDCG@$k$), and Mean Reciprocal Rank (MRR@$k$), at cutoffs $k \in {5,10,20}$. Recall@$k$ measures the fraction of relevant items retrieved in the top-$k$ list, while NDCG@$k$ accounts for ranking position by assigning higher scores to relevant items appearing earlier. MRR@$k$ measures the position of the first relevant item in the ranked list. Higher values indicate better performance for all three metrics.
\subsection{Baselines}
\label{sec:baselines}
To showcase the superiority of the proposed approach, we compare it against the following competitive baselines spanning RNN-based, MLP-based, attention-based, contrastive, linear, and state-space approaches:
 
\begin{itemize}
    \item \textbf{GRU4Rec}~\cite{hidasi2015session}: Session-based recommendation using gated recurrent units, modeling sequential patterns via hidden state transitions.
    \item \textbf{BERT4Rec}~\cite{sun2019bert4rec}: Bidirectional transformer for sequential recommendation with masked item prediction pre-training.
    \item \textbf{LightSANs}~\cite{fan2021lighter}: Low-rank decomposed self-attention to reduce the computational cost of next-item prediction.
    \item \textbf{FMLPRec}~\cite{zhou2022filter}: All-MLP model using frequency-domain filtering via fast fourier transform as a drop-in replacement for self-attention.
    \item \textbf{LinRec}~\cite{liu2023linrec}: Linear attention mechanism reducing self-attention complexity from $\mathcal{O}(T^2)$ to $\mathcal{O}(T)$.
    \item \textbf{BSARec}~\cite{shin2024attentive}: Fourier-based attentive inductive bias to mitigate oversmoothing in deep self-attention.
    \item \textbf{LRURec}~\cite{yue2024linear}: Parallelizable linear recurrent units for stable long-range dependency modeling.
    \item \textbf{GLINT-RU}~\cite{zhang2025glint}: Gated lightweight intelligent recurrent units for efficient sequential recommendation.
    \item \textbf{MUFFIN}~\cite{baek2025muffin}: Mixture of user-adaptive frequency filtering for sequential recommendation.
    \item \textbf{SIGMA}~\cite{liu2025sigma}: Selective gated Mamba for sequential recommendation.
\end{itemize}
 
% \subsection{Evaluation Metrics}
% \label{sec:evalMetrics}
% To compare the effectiveness of each algorithm, we evaluate recommendation quality using three standard ranking metrics, Recall, normalized discounted cumulative gain (NDCG), and mean reciprocal rank (MRR) at cutoffs $k \in \{5, 10, 20\}$.

% Recall@$k$ measures the fraction of relevant items retrieved in the top-$k$ list, averaged over all test users $\mathcal{U}_t$:
% \begin{equation}
%     \mathrm{Recall}@k = \frac{1}{|\mathcal{U}_t|}\sum_{u \in \mathcal{U}_t}\frac{|\mathrm{rel}_u(k)|}{|\mathrm{rel}_u|}.
% \end{equation} 
% where $\mathrm{rel}_u$ is the set of actual relevant items (ground-truth) for user $u$ and $\mathrm{rel}_u(k)$ is the set of relevant items for user $u$ in the provided top-$k$ suggestions. 

% NDCG@$k$ evaluates ranking quality by rewarding higher positions for relevant items:
% \begin{equation}
%     \mathrm{NDCG}@k = \frac{1}{|\mathcal{U}_t|}\sum_{u \in \mathcal{U}_t}
%     \frac{\sum_{i=1}^{k} s_i^u / \log_2(i+1)}{\sum_{i=1}^{|\mathrm{rel}_u|} 1/\log_2(i+1)},
% \end{equation}
% where $s_i^u \in \{0,1\}$ is the binary relevance of the $i$-th ranked item for user $u$.
 
% MRR@$k$ measures how early the first relevant item appears in the top-$k$ suggestions:
% \begin{equation}
%     \mathrm{MRR}@k = \frac{1}{|\mathcal{U}_t|}\sum_{u \in \mathcal{U}_t}\frac{1}{\mathrm{rank}_u},
% \end{equation}
% where $\mathrm{rank}_u$ is the position of the first relevant item in the top-$k$ list.

% Each evaluation metric yields values in the range $[0,1]$, where a score of $1$ indicates the best recommendation performance.

\subsection{Parameter Configuration}
\label{sec:paramConfig}
To ensure a fair comparison, we perform extensive hyperparameter tuning for all competing methods. All models are implemented in PyTorch. The embedding dimension is fixed at $d=64$, consistent across all experiments. For \textit{MoR4Rec}, the transformer backbone uses $L=4$ unrolled layers with $R=2$ recursion steps and $H=2$ attention heads with grouped query attention (2 key-value heads). The router is a single linear layer with softmax activation. We employ the \textit{middle-cycle} KV sharing strategy with cache update enabled. The gating variant ($G_1$: vanilla, headwise, or elementwise) is selected by validation performance. The hyperparameters for baseline approaches are tuned within the ranges reported in their respective papers. For all comparing algorithms, dropout and attention dropout rates are searched in $\{0.1, 0.2, 0.3, 0.4, 0.5\}$. 

Models are optimized with AdamW (learning rate $10^{-3}$, weight decay $0.01$) with a batch size of $256$, for up to $200$ epochs. Early stopping is applied based on validation $\mathrm{Recall}@10$ (patience $= 10$). The optimal configuration is selected based on the average of $\mathrm{Recall}@10$ and $\mathrm{MRR}@10$ on the validation set.

\pagestyle{empty}
\begin{landscape}
\begin{table*}[ht!]
    \renewcommand{\arraystretch}{2}
    \centering
    \caption{Recommendation performance comparison on three datasets. Mean rank is shown alongside each score. The best baseline result is \underline{underlined}. Imp.(\%) is the relative improvement of \textit{MoR4Rec} over the best baseline.}
    \adjustbox{width=\linewidth}{
    \begin{tabular}{llccccccccccccccccccccccc}
    \hline
    \textbf{Dataset} & \textbf{Metric} & \multicolumn{1}{c}{\textbf{GRU4Rec~\cite{hidasi2015session}}} & & \multicolumn{1}{c}{\textbf{BERT4Rec~\cite{sun2019bert4rec}}} & & \multicolumn{1}{c}{\textbf{LightSANs~\cite{fan2021lighter}}} & & \multicolumn{1}{c}{\textbf{FMLPRec~\cite{zhou2022filter}}} & & \multicolumn{1}{c}{\textbf{LinRec~\cite{liu2023linrec}}} & & \multicolumn{1}{c}{\textbf{BSARec~\cite{shin2024attentive}}} & & \multicolumn{1}{c}{\textbf{LRURec~\cite{yue2024linear}}} & & \multicolumn{1}{c}{\textbf{GLINT-RU~\cite{zhang2025glint}}} & & \multicolumn{1}{c}{\textbf{MUFFIN~\cite{baek2025muffin}}} & & \multicolumn{1}{c}{\textbf{SIGMA~\cite{liu2025sigma}}} & & \multicolumn{1}{c}{\textbf{\textit{MoR4Rec   (Proposed)}}} & & \textbf{Improv. (\%)} \\
    \hline
    \multirow{9}{*}{\textbf{MovieLens 1M}}
     & \textbf{Recall@5} & 0.2085 & 4 & 0.1274 & 6 & 0.1075 & 7 & 0.0898 & 10 & 0.1443 & 5 & 0.2185 & 3 & {\ul 0.2268} & 2 & 0.0843 & 11 & 0.0963 & 8 & 0.0949 & 9 & \textbf{0.2377} & 1 & 4.8060 \\
     & \textbf{Recall@10} & 0.2941 & 4 & 0.2019 & 6 & 0.1622 & 7 & 0.1449 & 9 & 0.2223 & 5 & 0.3077 & 3 & {\ul 0.3134} & 2 & 0.1356 & 11 & 0.1459 & 8 & 0.1416 & 10 & \textbf{0.3219} & 1 & 2.7122 \\
     & \textbf{Recall@20} & 0.3939 & 4 & 0.2975 & 6 & 0.2415 & 7 & 0.2146 & 8 & 0.3251 & 5 & 0.4113 & 3 & {\ul 0.4129} & 2 & 0.2132 & 9 & 0.2103 & 10 & 0.2069 & 11 & \textbf{0.4239} & 1 & 2.6641 \\
     & \textbf{NDCG@5} & 0.1434 & 4 & 0.0834 & 6 & 0.0708 & 7 & 0.0586 & 10 & 0.0953 & 5 & 0.1514 & 3 & {\ul 0.1581} & 2 & 0.0534 & 11 & 0.0652 & 8 & 0.0618 & 9 & \textbf{0.1658} & 1 & 4.8703 \\
     & \textbf{NDCG@10} & 0.1710 & 4 & 0.1076 & 6 & 0.0884 & 7 & 0.0763 & 10 & 0.1204 & 5 & 0.1800 & 3 & {\ul 0.1860} & 2 & 0.0698 & 11 & 0.0812 & 8 & 0.0769 & 9 & \textbf{0.1929} & 1 & 3.7097 \\
     & \textbf{NDCG@20} & 0.1963 & 4 & 0.1317 & 6 & 0.1084 & 7 & 0.0937 & 9 & 0.1463 & 5 & 0.2061 & 3 & {\ul 0.2110} & 2 & 0.0893 & 11 & 0.0975 & 8 & 0.0934 & 10 & \textbf{0.2186} & 1 & 3.6019 \\
    \hline
    \multirow{6}{*}{\textbf{Industrial \&  Scientific}}
     & \textbf{Recall@5} & 0.0235 & 4 & 0.0233 & 6 & 0.0177 & 9 & 0.0146 & 10 & 0.0234 & 5 & {\ul 0.0320} & 2 & 0.0305 & 3 & 0.0217 & 7 & 0.0214 & 8 & 0.0143 & 11 & \textbf{0.0321} & 1 & 0.3125 \\
     & \textbf{Recall@10} & 0.0376 & 4 & 0.0346 & 6 & 0.0257 & 9 & 0.0215 & 10 & 0.0358 & 5 & {\ul 0.0461} & 2 & 0.0448 & 3 & 0.0317 & 7 & 0.031 & 8 & 0.0213 & 11 & \textbf{0.0465} & 1 & 0.8677 \\
     & \textbf{Recall@20} & 0.0574 & 4 & 0.0501 & 6 & 0.0367 & 9 & 0.0304 & 11 & 0.0523 & 5 & {\ul 0.0663} & 2 & 0.0647 & 3 & 0.0460 & 7 & 0.0446 & 8 & 0.0309 & 10 & \textbf{0.0669} & 1 & 0.9050 \\
     & \textbf{NDCG@5} & 0.0151 & 8 & 0.0161 & 5 & 0.0125 & 9 & 0.0097 & 11 & 0.0163 & 4 & {\ul 0.0221} & 2 & 0.0211 & 3 & 0.0152 & 6 & 0.0152 & 7 & 0.0100 & 10 & \textbf{0.0223} & 1 & 0.9050 \\
     & \textbf{NDCG@10} & 0.0196 & 6 & 0.0197 & 5 & 0.0151 & 9 & 0.0119 & 11 & 0.0203 & 4 & {\ul 0.0267} & 2 & 0.0257 & 3 & 0.0184 & 7 & 0.0183 & 8 & 0.0122 & 10 & \textbf{0.0268} & 1 & 0.3745 \\
     & \textbf{NDCG@20} & 0.0245 & 4 & 0.0236 & 6 & 0.0179 & 9 & 0.0142 & 11 & 0.0244 & 5 & {\ul 0.0318} & 2 & 0.0307 & 3 & 0.022 & 7 & 0.0217 & 8 & 0.0146 & 10 & \textbf{0.0319} & 1 & 0.3145 \\
    \hline
    \multirow{6}{*}{\textbf{Video Games}}
     & \textbf{Recall@5} & 0.0582 & 4 & 0.0519 & 6 & 0.0471 & 8 & 0.0349 & 11 & 0.0481 & 7 & 0.0684 & 3 & {\ul 0.0700} & 2 & 0.0433 & 9 & 0.0528 & 5 & 0.0403 & 10 & \textbf{0.0712} & 1 & 1.7143 \\
     & \textbf{Recall@10} & 0.0881 & 4 & 0.0791 & 5 & 0.0691 & 8 & 0.0541 & 11 & 0.0725 & 7 & 0.0991 & 3 & {\ul 0.1036} & 2 & 0.0667 & 9 & 0.0776 & 6 & 0.0605 & 10 & \textbf{0.1046} & 1 & 0.9653 \\
     & \textbf{Recall@20} & 0.1286 & 4 & 0.117 & 5 & 0.0994 & 8 & 0.0808 & 11 & 0.1066 & 7 & 0.1402 & 3 & {\ul 0.1474} & 2 & 0.0982 & 9 & 0.1113 & 6 & 0.0876 & 10 & \textbf{0.1492} & 1 & 1.2211 \\
     & \textbf{NDCG@5} & 0.0384 & 4 & 0.0345 & 6 & 0.0321 & 8 & 0.0228 & 11 & 0.0323 & 7 & 0.0467 & 3 & {\ul 0.0474} & 2 & 0.0288 & 9 & 0.0357 & 5 & 0.0272 & 10 & \textbf{0.0482} & 1 & 1.6888 \\
     & \textbf{NDCG@10} & 0.0480 & 4 & 0.0433 & 6 & 0.0392 & 8 & 0.0290 & 11 & 0.0401 & 7 & 0.0566 & 3 & {\ul 0.0583} & 2 & 0.0363 & 9 & 0.0437 & 5 & 0.0337 & 10 & \textbf{0.0589} & 1 & 1.0292 \\
     & \textbf{NDCG@20} & 0.0582 & 4 & 0.0528 & 5 & 0.0468 & 8 & 0.0357 & 11 & 0.0486 & 7 & 0.0669 & 3 & {\ul 0.0693} & 2 & 0.0442 & 9 & 0.0521 & 6 & 0.0405 & 10 & \textbf{0.0702} & 1 & 1.2987 \\
    \hline
        \end{tabular}}
    \label{tab:results}
\end{table*}
\end{landscape}
\pagestyle{plain}

\section{Results and Discussion}
\label{sec:resultsDiscuss}
In this section, we present the experimental results and answer each of the research questions defined in Section \ref{sec:researchQues}.
\label{sec:rnd}
\subsection{Overall Performance (RQ1)} \label{sec:main_results}

\begin{figure*}[ht]
    \centering
    \includegraphics[width=0.95\textwidth, height = 6cm]{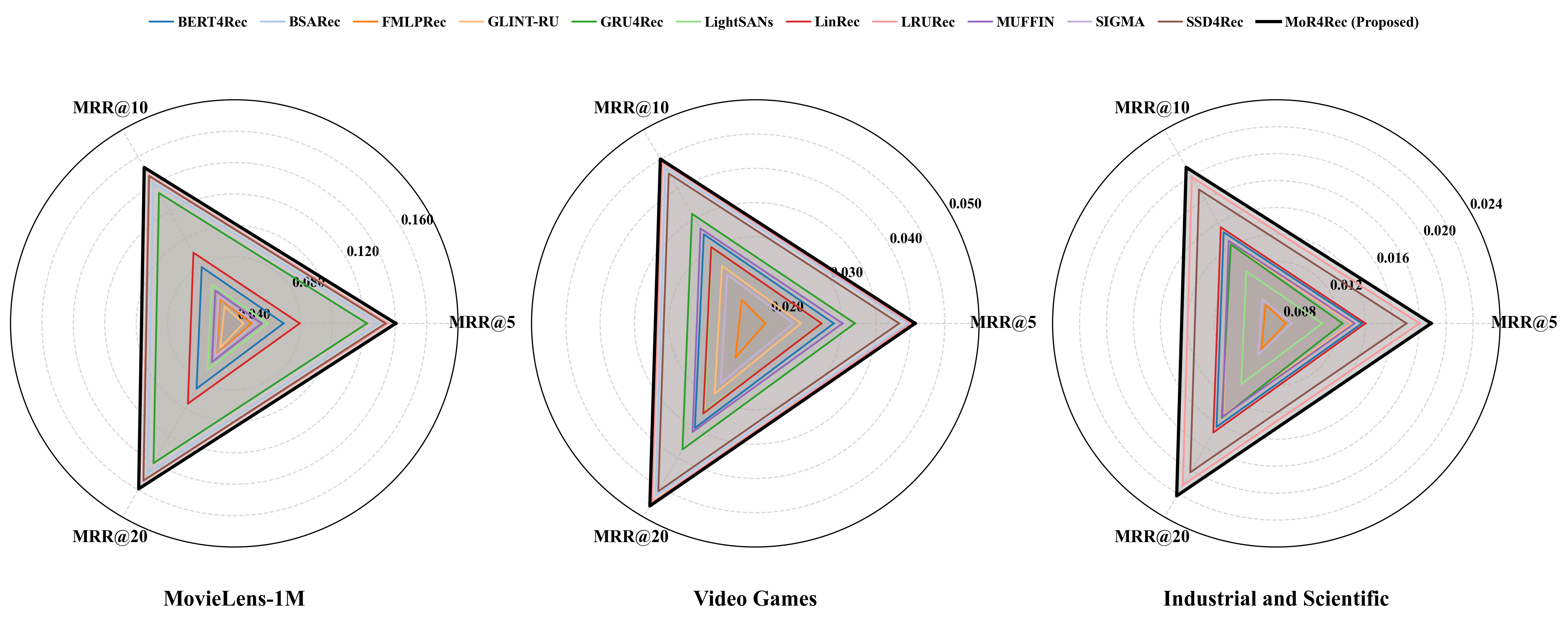}
    \caption{Performance of comparing algorithms in terms of $MRR@k$.}
    \label{fig:mrr}
\end{figure*}

Table~\ref{tab:results} compares \textit{MoR4Rec} with recent state-of-the-art baselines across three benchmark datasets. \textit{MoR4Rec} consistently achieves superior or competitive performance across all metrics, with gains at each cutoff $k$ demonstrating its ability to capture both precise top-ranked recommendations and broader relevant-item coverage. The consistent improvements, particularly in MRR@10 (Figure~\ref{fig:mrr}), indicate that \textit{MoR4Rec} ranks relevant items earlier in the recommendation list. As MRR rewards the position of the first relevant item, this advantage suggests that \textit{MoR4Rec} not only retrieves relevant items but places them higher in the list, which is particularly important in practical top-ranked recommendations. These gains support our central hypothesis that interactions require different computational depths: routine interactions can be processed shallowly, while preference shifts or long-range dependencies benefit from deeper recursive refinement. By dynamically allocating computation through routing, \textit{MoR4Rec} uses model capacity more effectively, improving representation quality and recommendation performance.

\subsection{Effect of Gated Attention Variants (RQ2)}
\label{sec:ablation}

% \begin{figure*}[ht]
%     \centering
%     \begin{subfigure}[b]{0.32\linewidth}
%         \centering
%         \includegraphics[width=\linewidth]{Hyper_1M.jpg}
%         \caption{MovieLens 1M}
%         \label{fig:hyper_1m}
%     \end{subfigure}
%     \hfill
%     \begin{subfigure}[b]{0.32\linewidth}
%         \centering
%         \includegraphics[width=\linewidth]{Hyper_IAS.jpg}
%         \caption{Industrial \& Scientific}
%         \label{fig:hyper_ias}
%     \end{subfigure}
%     \hfill
%     \begin{subfigure}[b]{0.32\linewidth}
%         \centering
%         \includegraphics[width=\linewidth]{Hyper_VG.jpg}
%         \caption{Video Games}
%         \label{fig:hyper_vg}
%     \end{subfigure}
%     \caption{Hyperparameter sensitivity of \textit{MoR4Rec} to dropout and attention dropout rates across three datasets. Higher (yellower) surface regions indicate better performance.}
%     \label{fig:sensitivity}
% \end{figure*}

\begin{figure*}[ht]
    \centering
    \includegraphics[ width=0.95\textwidth, height = 4cm]{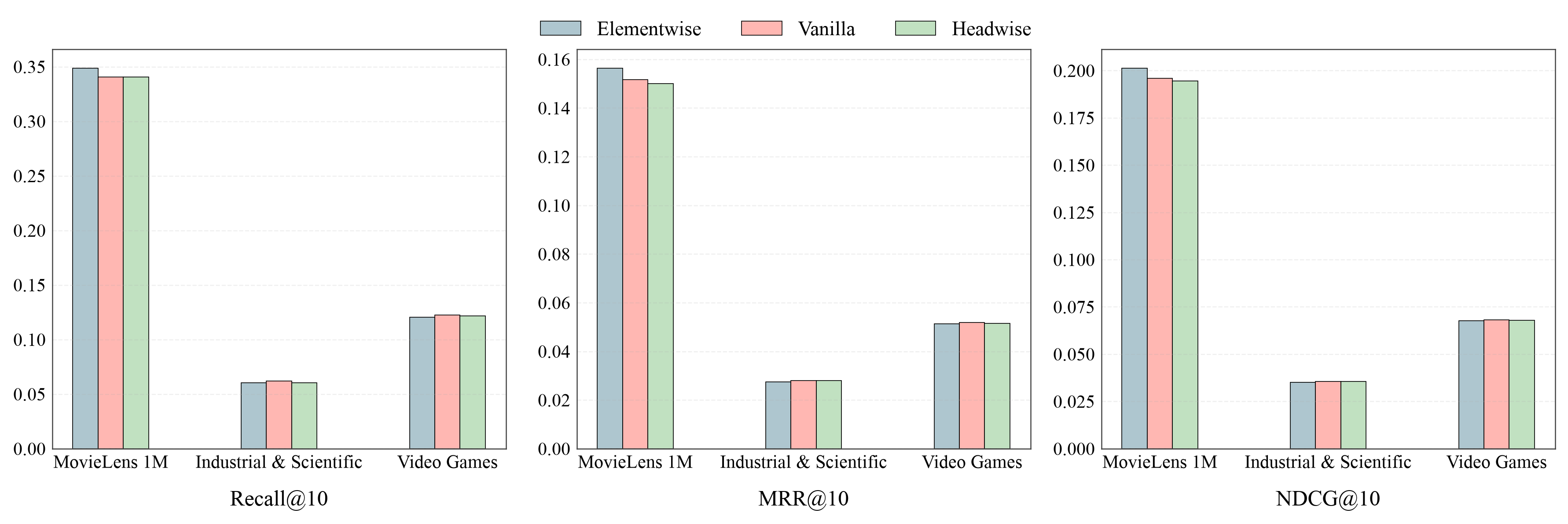}
  \caption{Ablation of $G_1$ gating variants across three datasets on Recall@10, MRR@10, and NDCG@10.}
    \label{fig:ablation_gating}
\end{figure*}

We compare the three gated attention strategies within the \textit{MoR4Rec} framework: (i) \textit{vanilla} (standard attention, no gating), (ii) \textit{headwise} (one sigmoid gate per attention head), and (iii) \textit{elementwise} (one sigmoid gate per feature dimension). Figure~\ref{fig:ablation_gating} reports the results in terms of Recall@10, MRR@10, and NDCG@10 across all three datasets. On denser datasets, elementwise gating ($G_1^{\text{ew}}$) achieves a consistent marginal advantage, as fine-grained feature-level modulation better exploits the richer interaction context available. On sparser datasets, vanilla attention proves more competitive, suggesting that additional gating parameters risk overfitting to noisy interaction patterns in low-signal regimes. Headwise gating ($G_1^{\text{hw}}$) occupies a middle ground in both settings. Overall, elementwise gating is preferred for interaction-rich datasets, while vanilla attention provides a stronger default under data scarcity.

\subsection{Per-Token Recursion Depth Analysis (RQ3)}

\begin{figure}[h]
    \centering
    \includegraphics[width=0.85\columnwidth]{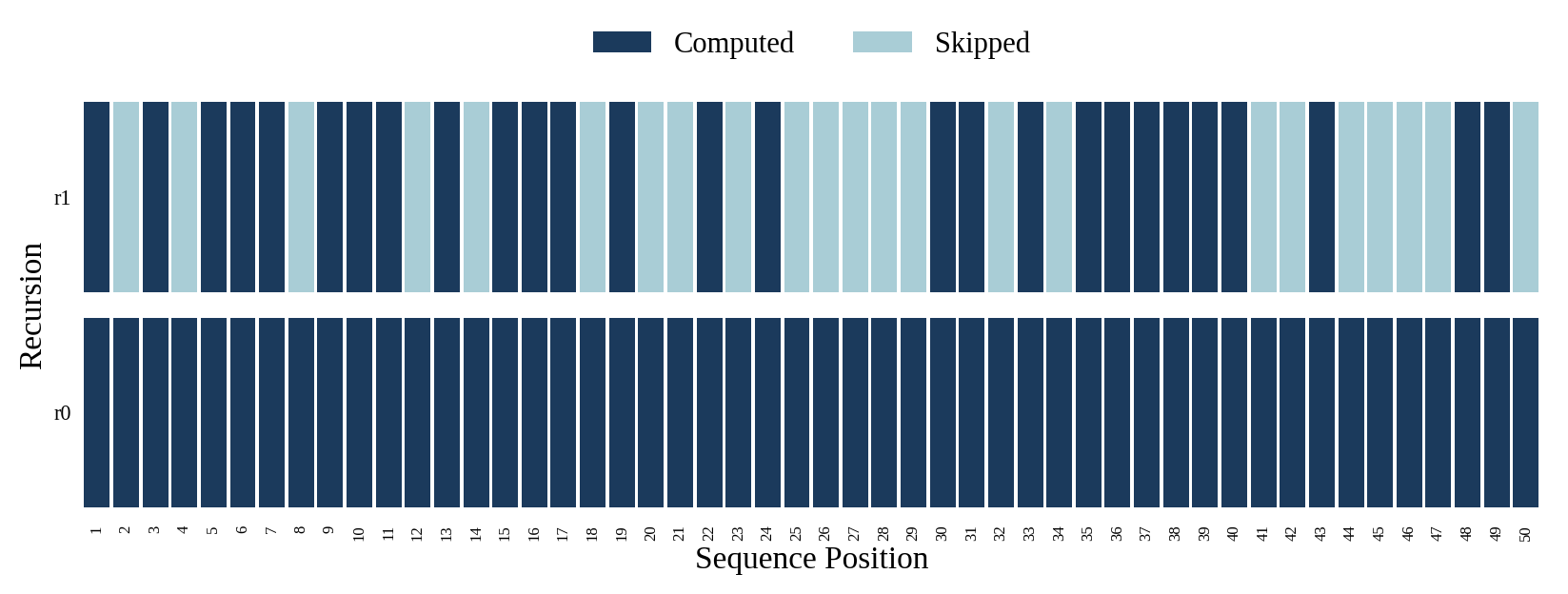}
    \caption{Per-token recursion depth for a random user with $R=2$.}
    \label{fig:routing_r2}
\end{figure}

Figure~\ref{fig:routing_r2} shows the recursion depth the router picks for
each token. We trace one random user sequence at test time under a budget of
$R=2$. A few things stand out. Every token is computed at $r_0$. So each
position recurses at least once. The number of computed tokens then drops at
$r_1$. A token routed to depth $k$ is recursed at levels $r_0$ to $r_{k-1}$
and skipped after that. This gives the nested shape we see in the figure. The
choice is also token specific. Some tokens exit after the first recursion.
Others are carried to the deeper level. This tells us the router learns where
to spend compute. It does not apply a fixed depth to the whole sequence. We
also see recent positions kept at the deeper level more often. That fits the
idea that recent items matter more for the next-item prediction.
\label{sec:recursion}

\subsection{Does Adaptive Recursion Reduce Computational Cost? (RQ4)}
\label{sec:flops}

We compare MoR4Rec with a vanilla transformer encoder for sequential recommendation. Both models use four layers ($L=4$) and the same scoring term, providing a controlled setting to isolate the backbone effect. As shown in Table~\ref{tab:flops}, MoR4Rec remains more computationally efficient despite not being a stripped-down encoder. Its recursive block includes a sigmoid gating mechanism that adds some compute to the shared weights. Even with this gate, MoR4Rec reduces computational and parameter costs across all three datasets. 
\begin{table}[ht]
\centering
\caption{Computational cost across datasets. MFLOPs, parameters, training time, and inference time are reported per user forward pass.}
\label{tab:flops}
\footnotesize
\setlength{\tabcolsep}{4pt}
\renewcommand{\arraystretch}{0.8}
\resizebox{0.8\columnwidth}{!}{%
\begin{tabular}{l|lrrrr}
\toprule
Dataset & Model  & MFLOPs & Parameter & Train & Infer \\
\midrule
\multirow{2}{*}{MovieLens 1M}&Vanilla Transformer & 44.10 &  422.0 & 647.1 & 0.0004 \\
&MoR4Rec    &  42.76 & 380.5 & 643.7 & 0.0003 \\ \midrule
\multirow{2}{*}{Industrial \&  Scientific}&Vanilla Transformer & 187.66 & 1857.66 & 897.26 & 0.0002 \\
&MoR4Rec  &  184.48 & 1803.84 & 1070.64 & 0.0002 \\ \midrule
\multirow{2}{*}{Video Games}&Vanilla Transformer & 186.15 & 1842.56 & 2339.53 & 0.0002 \\
&MoR4Rec    &  182.78  & 1788.86  & 2302.33 & 0.0002 \\
\bottomrule
\end{tabular}%
}
\end{table}
On MovieLens 1M, it requires $42.76$ MFLOPs versus $44.10$ MFLOPs and uses $380.5$K parameters versus $422.0$K for the vanilla transformer. Similar reductions are observed on Industrial \& Scientific and Video Games. Training time is lower for MoR4Rec on MovieLens 1M and Video Games, while it is higher on Industrial \& Scientific. Inference is faster on MovieLens 1M and comparable on the other two datasets. Overall, under a matched architecture and scoring budget, MoR4Rec is lighter in computation and parameter count, while maintaining comparable inference efficiency. These results indicate that adaptive per-token recursion uses the computational budget more effectively than a fixed-depth encoder.

% \subsection{Hyperparameter Sensitivity (RQ4)}
% \label{sec:sensitivity}

% We study the sensitivity of \textit{MoR4Rec} to the two primary regularization hyperparameters: \textbf{dropout rate} $p \in \{0.1, 0.2, 0.3, 0.4, 0.5\}$ applied to item embeddings and FFN outputs, and \textbf{attention dropout rate} $p_{\mathrm{attn}} \in \{0.1, 0.2, 0.3, 0.4, 0.5\}$ applied to the attention weight matrix. Figure~\ref{fig:sensitivity} shows 3D surface plots of the combined score (average of Recall@10 and MRR@10) over the joint grid of both hyperparameters for each dataset.
% The surface plots reveal distinct dataset-dependent regularization patterns. On MovieLens 1M, performance improves monotonically as both dropout and attention dropout increase toward $\{0.1, 0.4\}$. This behavior is characteristic of dense datasets as the rich interaction signal makes the model susceptible to over-reliance on specific patterns, and stronger dropout regularization promotes better generalization. On sparse datasets, the pattern is markedly different. Both sparse datasets prefer low attention dropout ($p_{\mathrm{attn}} \in \{0.5, 0.1\}$) combined with moderate dropout ($p \in \{0.5, 0.2\}$). This finding motivates tuning attention dropout independently of standard dropout, as their optimal values diverge across interaction density regimes. Overall, \textit{MoR4Rec} maintains strong performance across a broad range of configurations. The performance variation across the entire grid is less than $15\%$ relative to all datasets, confirming its practical robustness to hyperparameter choice. 

\subsection{Statistical Significance Analysis (RQ5)}
\label{sec:stats}
 
To rigorously validate observed performance gains, we conduct nonparametric statistical significance testing following the protocol of Dem\v{s}ar~\cite{demvsar2006statistical}. We apply the Friedman test over $\mathcal{K}=11$ competing methods and $\mathcal{N}=3$ datasets, testing whether all models perform equivalently. Table~\ref{tab:statistical} reports the Friedman statistic $F_F$ alongside the critical value at $\alpha = 0.05$.
 
\begin{table}[h]
\renewcommand{\arraystretch}{0.8}
\centering
\small
\caption{Friedman and Nemenyi statistical analysis summary.}
\adjustbox{max width=\linewidth}{
    \begin{tabular}{lccccccc}
    \toprule
    \textbf{Metric} & $F_F$ & \textbf{Crit. Val.} $(\alpha{=}0.05)$ & $q_\alpha$ & $\mathcal{N}$ & $\mathcal{K}$ & \textbf{CD} \\
    \midrule
    \textbf{Recall@5} & 27.1177 & \multirow{3}{*}{2.3479} & \multirow{3}{*}{3.2190} & \multirow{3}{*}{3} & \multirow{3}{*}{11} & \multirow{3}{*}{8.7171} \\
    \textbf{NDCG@5}   & 15.3684 & & & & & \\
    \textbf{MRR@5}    & 24.4000 & & & & & \\
    \bottomrule
    \end{tabular}
    }
    \label{tab:statistical}
\end{table}

 \begin{figure*}[t]
    \centering
    \begin{subfigure}[b]{0.48\linewidth}
        \centering
        \includegraphics[width=\linewidth, trim={0 20 0 72}, clip]{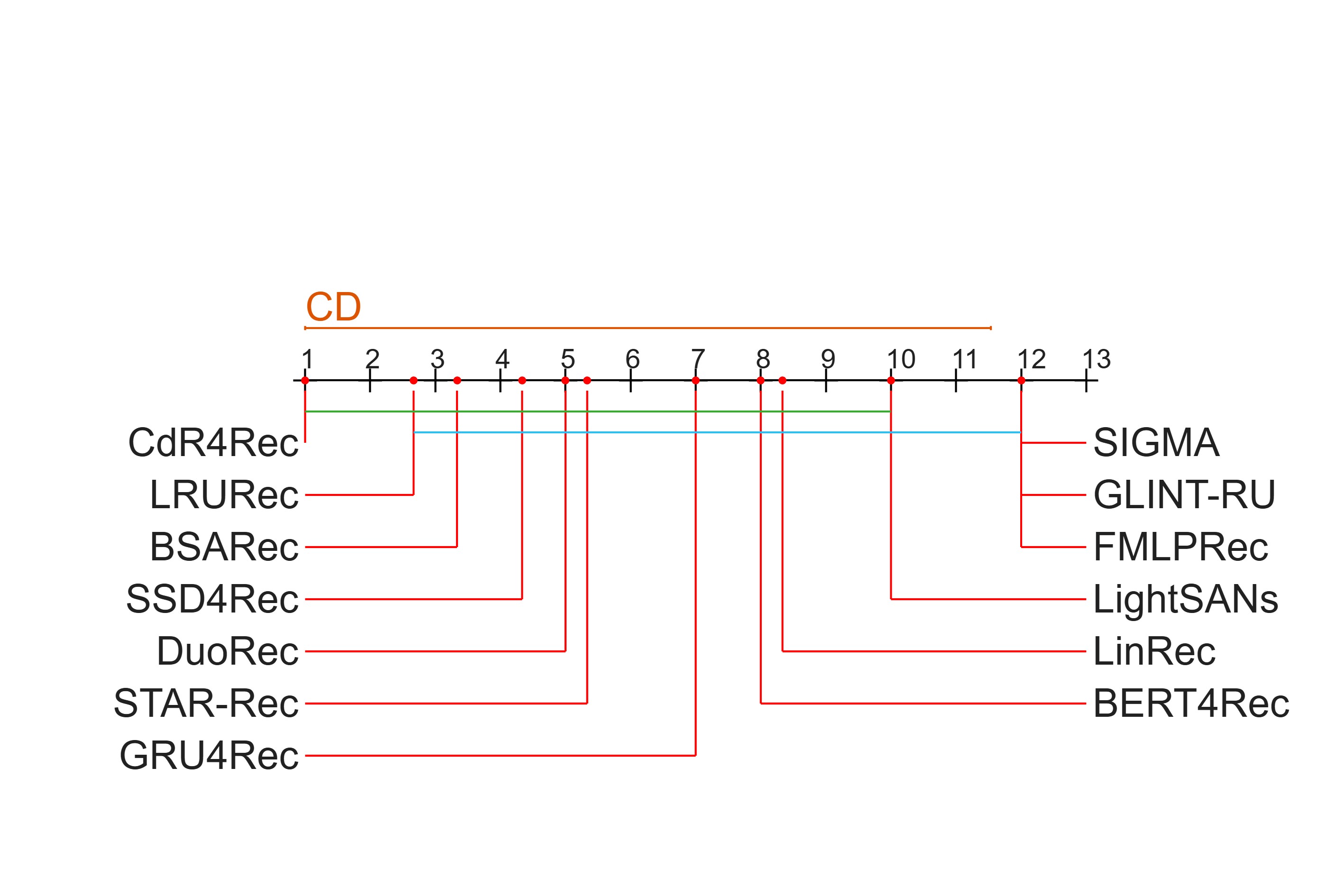}
        \caption{$\mathrm{Recall}@5$}
        \label{fig:r5_nemenyi}
    \end{subfigure}
    \hfill
    \begin{subfigure}[b]{0.48\linewidth}
        \centering
        \includegraphics[width=\linewidth, trim={0 20 0 72}, clip]{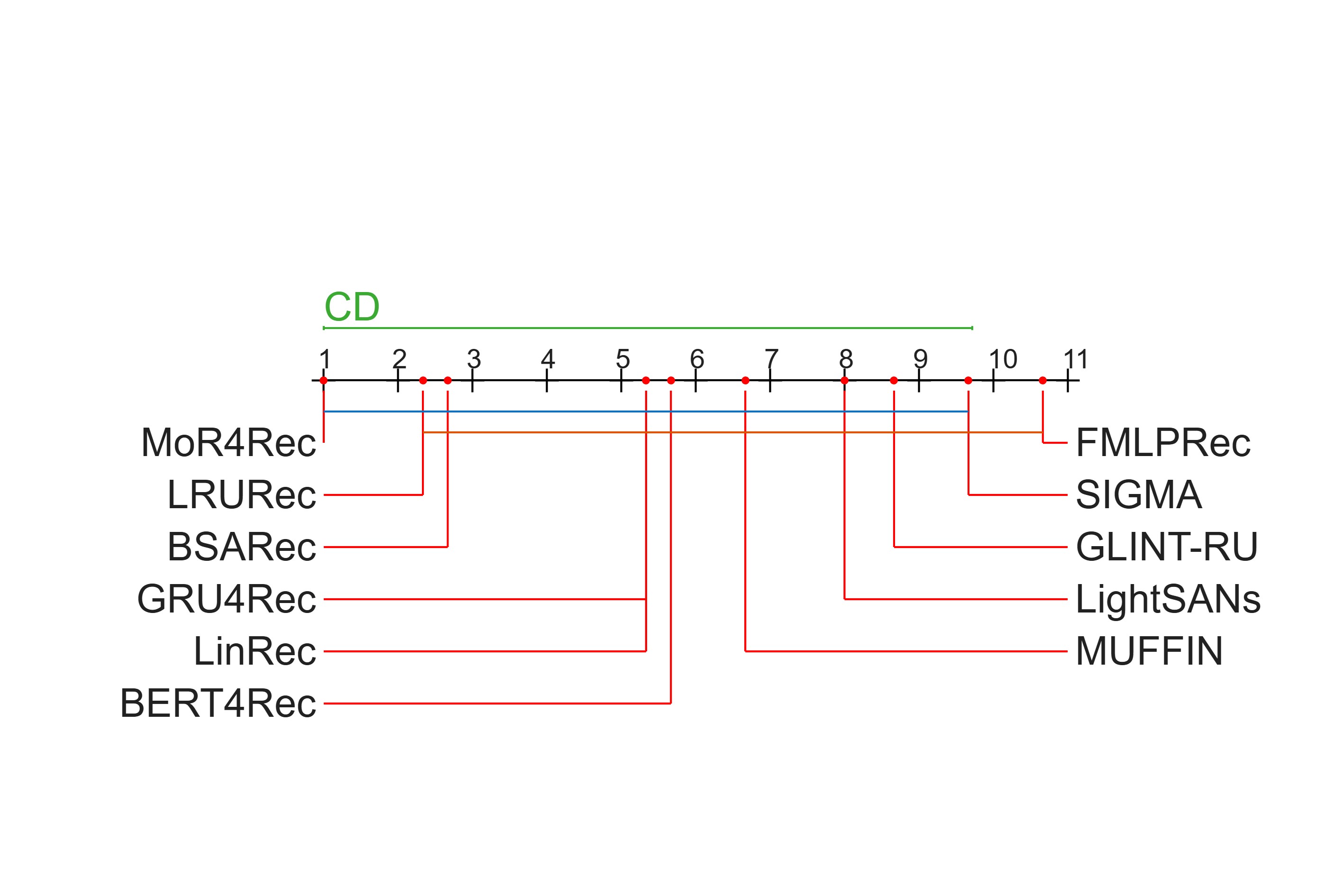}
        \caption{$\mathrm{NDCG}@5$}
        \label{fig:n5_nemenyi}
    \end{subfigure}

    \vspace{2mm}
    \begin{subfigure}[b]{0.48\linewidth}
        \centering
        \includegraphics[width=\linewidth, trim={0 20 0 72}, clip]{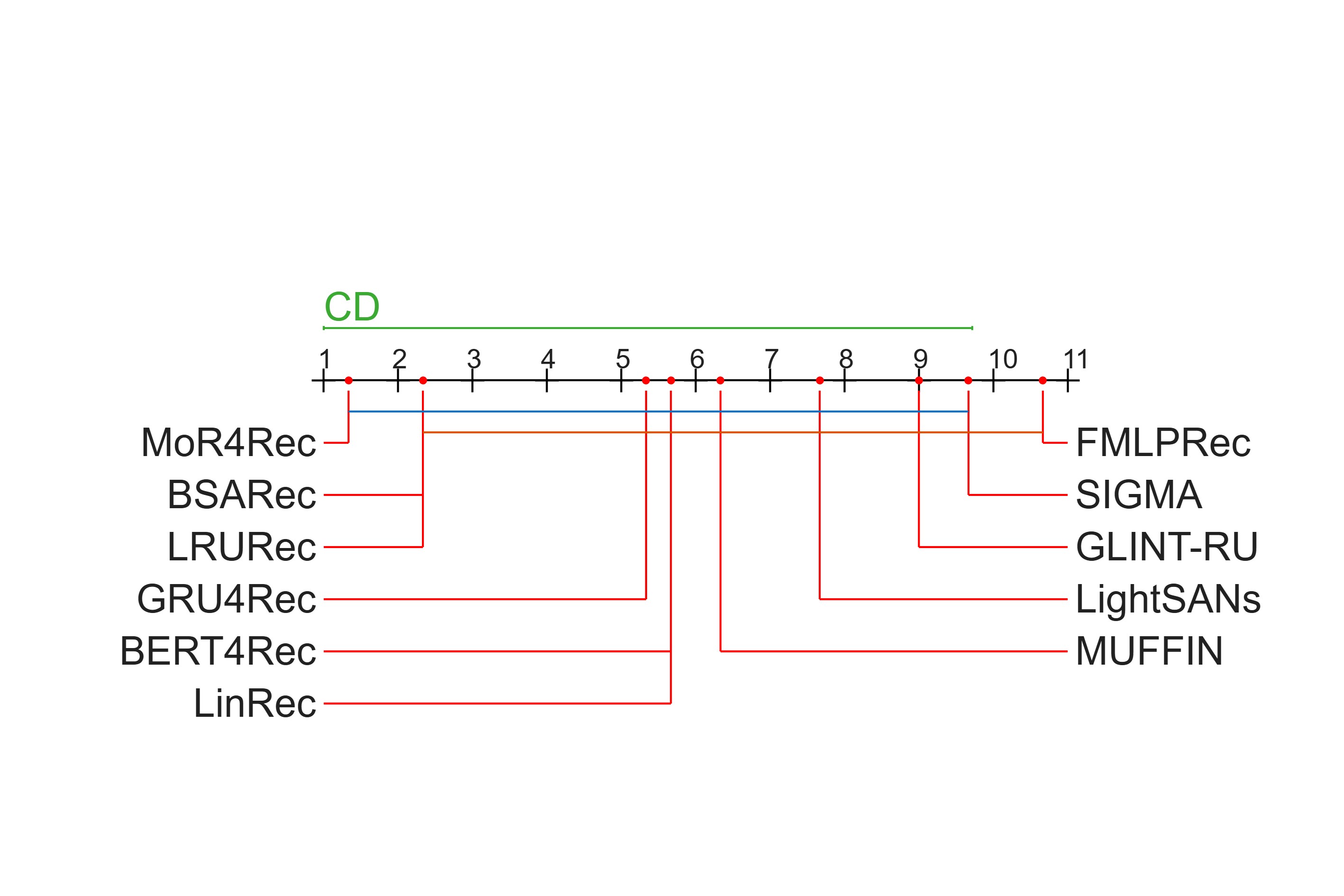}
        \caption{$\mathrm{MRR}@5$}
        \label{fig:m5_nemenyi}
    \end{subfigure}
    \caption{Critical difference (CD) diagrams for Recall@5, NDCG@5, and MRR@5.}
    \label{fig:cd_diagram}
\end{figure*}
For all three metrics, $F_F$ substantially exceeds the critical value, leading to rejection of the null hypothesis that all models perform equivalently ($p < 0.05$). We then apply the post-hoc Nemenyi test to perform pairwise comparisons. The critical difference $CD$ is $CD = q_\alpha\sqrt{\frac{\mathcal{K}(\mathcal{K}+1)}{6\mathcal{N}}}$, where $q_\alpha = 3.2190$ for $\mathcal{K}=11$ at $\alpha=0.05$. As illustrated in Figure~\ref{fig:cd_diagram}, \textit{MoR4Rec} achieves a statistically superior average rank compared to all competing baselines, with rank differences exceeding $CD = 8.7171$, confirming the robustness and consistency of its performance improvements.
 
\section{Conclusion}
\label{sec:conclusion}
% ============================================================

We presented \textit{MoR4Rec}, a sequential recommendation framework that introduces interaction-level adaptive computation through a Mixture-of-Recursion paradigm. Motivated by the observation that interactions within a user sequence may require different levels of contextual reasoning, \textit{MoR4Rec} dynamically allocates computation across recursion depths while maintaining efficiency through shared recursive modeling and key--value reuse. A gated attention mechanism further improves the modeling of contextual dependencies, while auxiliary routing objectives support stable training. Experiments on three benchmark datasets against fifteen competitive baselines show that \textit{MoR4Rec} consistently achieves strong performance and favorable efficiency--accuracy trade-offs, with ablation studies further validating the contribution of its core components.

These results highlight the promise of adaptive recursion for sequential recommendation. Future work includes improving the interpretability of learned routing behavior and extending the framework to more challenging settings such as multi-modal and cross-domain recommendation settings, where behavioral heterogeneity is even more pronounced.

\clearpage
%\newpage
\bibliographystyle{acm}
\bibliography{ACMbiblo}
\end{document}